\documentclass[prb,preprint,superscriptaddress]{revtex4}

\usepackage{amsmath,amsfonts,amssymb}
\usepackage{graphicx}
\usepackage[english]{babel}
\usepackage{color}
\usepackage{booktabs,longtable}

\begin{document}

\author{J.-C. Rojas-S\'anchez}
\affiliation{INAC/SP2M, CEA-UJF, F-38054 Grenoble, France}

\author{M. Cubukcu}
\affiliation{INAC/SP2M, CEA-UJF, F-38054 Grenoble, France}

\author{A. Jain}
\affiliation{INAC/SP2M, CEA-UJF, F-38054 Grenoble, France}

\author{C. Vergnaud}
\affiliation{INAC/SP2M, CEA-UJF, F-38054 Grenoble, France}

\author{C. Portemont}
\affiliation{CROCUS-Technology, F-38025 Grenoble, France}

\author{C. Ducruet}
\affiliation{CROCUS-Technology, F-38025 Grenoble, France}

\author{A. Barski}
\affiliation{INAC/SP2M, CEA-UJF, F-38054 Grenoble, France}

\author{A. Marty}
\affiliation{INAC/SP2M, CEA-UJF, F-38054 Grenoble, France}

\author{L. Vila}
\affiliation{INAC/SP2M, CEA-UJF, F-38054 Grenoble, France}

\author{J.-P. Attan\'e}
\affiliation{INAC/SP2M, CEA-UJF, F-38054 Grenoble, France}

\author{E. Augendre}
\affiliation{LETI, CEA, Minatec Campus, F-38054 Grenoble, France}

\author{G. Desfonds}
\affiliation{INAC/SCIB, CEA-UJF, F-38054 Grenoble, France}

\author{S. Gambarelli}
\affiliation{INAC/SCIB, CEA-UJF, F-38054 Grenoble, France}

\author{H. Jaffr\`es}
\affiliation{UMP CNRS-Thal\`es, CNRS, F-91767 Palaiseau, France}

\author{J.-M. George}
\affiliation{UMP CNRS-Thal\`es, CNRS, F-91767 Palaiseau, France}

\author{M. Jamet}
\affiliation{INAC/SP2M, CEA-UJF, F-38054 Grenoble, France}

\title{Spin Pumping and Inverse Spin Hall Effect in Germanium}

\date{\today}

\begin{abstract}
We have measured the inverse spin Hall effect (ISHE) in
\textit{n}-Ge at room temperature. The spin current in germanium was
generated by spin pumping from a CoFeB/MgO magnetic tunnel junction in order to prevent the impedance mismatch issue.
A clear electromotive force was measured in Ge at the ferromagnetic
resonance of CoFeB. The same study was then carried out on several
test samples, in particular we have investigated the influence of
the MgO tunnel barrier and sample annealing on the ISHE signal.
First, the reference CoFeB/MgO bilayer grown on SiO$_{2}$ exhibits a
clear electromotive force due to anisotropic magnetoresistance and
anomalous Hall effect which is dominated by an asymmetric
contribution with respect to the resonance field. We also found that
the MgO tunnel barrier is essential to observe ISHE in Ge and that
sample annealing systematically lead to an increase of the signal.
We propose a theoretical model based on the presence of localized
states at the interface between the MgO tunnel barrier and Ge to
account for these observations. Finally, all of our results are
fully consistent with the observation of ISHE in heavily doped
$n$-Ge and we could estimate the spin Hall angle at room temperature
to be $\approx$0.001.
\end{abstract}


\maketitle

\section{Introduction}
The first challenging requirement to develop semiconductor (SC)
spintronics\cite{Zutic2004,Awschalom2007} \textit{i.e.} using both carrier charge and spin in electronic
devices consists in injecting spin polarized electrons in the
conduction band of a SC at room temperature. SCs should be further
compatible with silicon mainstream technology for implementation in
microelectronics making silicon, germanium and their alloys among
the best candidates.\cite{Zutic2011} In Si, due to low spin-orbit coupling, very
long spin diffusion lengths were predicted and measured
experimentally.\cite{Dash2009,Jonker2007,Appelbaum2007,Suzuki2011} Germanium exhibits the same crystal inversion
symmetry as Si, a low concentration of nuclear spins but higher carrier
mobility and larger spin-orbit coupling which should allow in
principle spin manipulation by electric fields such as the Rashba
field.\cite{Zhou2011,Jain2011,Saito2011,Jeon2011,Hanbicki2012} So far in order to perform spin injection from a
ferromagnetic metal (FM) into Si or Ge, one needs to overcome at least
three major obstacles: \textit{(i)} the conductivity mismatch which requires
the use of a highly-resistive spin-conserving interface between the
FM and the SC,\cite{Fert2001} \textit{(ii)} the Fermi level pinning at the SC surface due
to the presence of a high density of interface states and the
interface spin flips which are generally associated\cite{Dash2009,Jain2011,Tran2009} and
finally \textit{(iii)} the presence of random magnetic stray fields created by
surface magnetic charges at rough interface\cite{Jain2011,Dash2011} around which the electrically
injected spins are precessing and partly lost by decoherence. In
this work, we have inserted a thin MgO tunnel barrier between Ge and
the CoFeB ferromagnetic electrode in order to: \textit{(i)} circumvent the
conductivity mismatch and \textit{(ii)} partly alleviate Fermi level pinning
by strongly reducing the interface states density\cite{Cantoni2011,Lee2010,Zhou2010} which leads
to a modest Schottky barrier height at the MgO/\textit{n}-Ge interface. We
have then investigated the spin injection mechanisms using the so called
three-terminal device.\cite{Dash2009} In this geometry, the same ferromagnetic
electrode is used for spin injection and detection. This three-terminal device used
in non-local geometry represents a simple and unique tool to probe spin
accumulation both into interface states and in the SC channel.\cite{Dash2009,Jain2011,Li2011} In particular, we could measure spin injection in the silicon and germanium conduction bands at room temperature.\cite{Jain2012,Jain2013}\\
The spin Hall angle ($\theta_{SHE}$, ratio between the transverse
spin current density and the longitudinal charge current density)
\cite{Vila2007} is a key material parameter to develop new kinds of devices based on the spin Hall effect (SHE). The SHE is the conversion of a charge current into a spin current via the spin-orbit interaction. Conversely the inverse spin Hall effect (ISHE) is the conversion of a spin current into a charge current. Several methods have been developed to determine quantitatively $\theta_{SHE}$: pure
magnetotransport measurements on lateral spin valves (LSV),\cite{Vila2007,Valenzuela2006,Morota2011} ferromagnetic
resonance (FMR) along with spin pumping (SP-FMR)
\cite{Ando2011a,Mosendz2010b,Azevedo2011} and spin torque FMR
(ST-FMR) \cite{Liu2011} on ferromagnetic/non-magnetic bilayers
(FM/N). Recently $\theta_{SHE}$ could be estimated in $n$ and $p$-GaAs \cite{Ando2011b} as well as in $p$-Si \cite{Shikoh2013} by spin pumping and inverse spin Hall effect. The precession of the FM layer in direct contact with the SC has been excited by microwaves which pumps a spin current into the SC. The spin current was then detected by inverse spin Hall effect. In that case, the interface resistance to overcome the conductivity mismatch issue was given by the reminiscent Schottky barrier at the FM/SC interface. Here we have similarly used a combined SP-FMR method to study inverse spin Hall effect in germanium. In the first section, we describe the sample preparation and the experimental techniques. In the second section, the phenomenological models for the ferromagnetic resonance and electromotive force are presented. Finally the experimental results are shown and discussed in sections 3 and 4 respectively. In particular, we propose a microscopic model based on the presence of localized states at the MgO/Ge interface to explain the spin pumping mechanism in our system. We finally discuss about the influence of the MgO tunnel barrier and sample annealing on the ISHE signal.

\section{Sample preparation}

\begin{figure}[h!]
\includegraphics[width=15 cm]{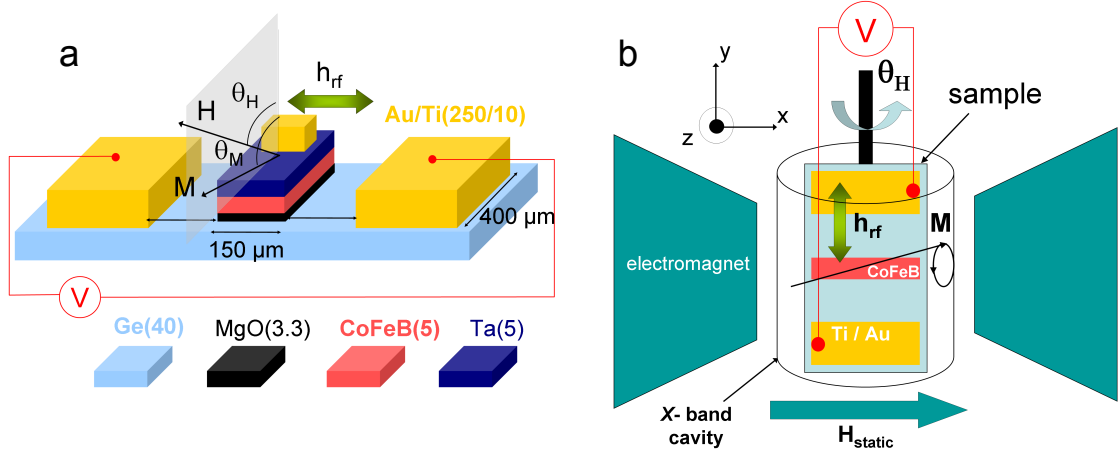}
\caption{(Color online) (a) Schematic drawing of the
multiterminal device used for SP-FMR measurements along with the definition of $\theta_H$
and $\theta_M$. $h_{rf}$ is the radiofrequency magnetic field. The thickness of each layer is given in nanometers between parenthesis. (b) Drawing of the device inserted into the cylindrical \textit{X}-band electron paramagnetic resonance (EPR) cavity.}
\label{fig01ab}
\end{figure}

The multi-terminal device we initially used for electrical spin
injection, detection and manipulation\cite{Jain2012} is made of a
full stack Ta(5 nm)/CoFeB(5 nm)/MgO(3 nm) grown by
sputtering on a 40 nm-thick germanium film on insulator (GOI).\cite{Jain2011} GOI wafers are
made of a Si $p$+ degenerate substrate and a 100 nm-thick SiO$_2$ layer (BOX).
They were fabricated using the Smart Cut$^{TM}$ process and Ge
epitaxial wafers.\cite{Deguet2005} The transferred 40 nm-thick Ge film was $n$-type
doped in two steps: a first step (phosphorus, $3\times10^{13}$
cm$^{-2}$, 40 keV, annealed for 1h at 550$^\circ$C) that provided
uniform doping in the range of 10$^{18}$ cm$^{-3}$, and a second
step (phosphorus, $2\times10^{14}$ cm$^{-2}$, 3 keV, annealed for 10
s at 550$^\circ$C) that increased surface $n$+ doping to the vicinity
of 10$^{19}$ cm$^{-3}$. The thickness of the \textit{n}+-doped layer
is estimated to be 10 nm. The GOI surface was finally capped with
amorphous SiO$_2$ to prevent Ge from surface oxidation. The tunnel barrier
and ferromagnetic electrode were then fabricated from magnesium (Mg,
1.1nm) and cobalt-iron-boron (Co$_{60}$Fe$_{20}$B$_{20}$, 5nm)
layers deposited by conventional DC magnetron sputtering onto
germanium (Ge) after removing the SiO$_2$ capping layer using
hydrofluoric acid and de-ionized water. The deposition rates were
respectively of 0.02 nm.s$^{-1}$ and 0.03 nm.s$^{-1}$ at an argon
pressure of $2\times10^{-3}$ mbar. The base pressure was
$7\times10^{-9}$ mbar. All the depositions were performed at room
temperature. The oxidation of the insulating barrier was performed by
plasma oxidation, exposing the Mg metallic layer to a 30 seconds
radio-frequency oxygen plasma at a pressure of $6\times10^{-3}$ mbar
and a radio-frequency power of 100 W. Three successive Mg deposition
plus oxidation steps were achieved ([Mg 1.1 / oxidation]$_3$) to
grow a 3.3 nm thick MgO layer. The sample annealings were performed under 10$^{-7}$
mbar at 300$^\circ$C for 90 minutes. The ferromagnetic layer is then
capped with 5 nm of Ta to prevent oxidation. After depositing the
spin injector, samples have been processed using standard optical
lithography. In a first step, we define the ferromagnetic electrode
($150\times400$ $\mu$m$^2$) and ohmic contacts ($300\times400$
$\mu$m$^2$ on Ge and $100\times100$ $\mu$m$^2$ on top of the
ferromagnetic electrode). In a second step, the germanium channel is
etched down to the BOX to form a mesa of 1070 $\mu$m long and 420
$\mu$m large. Finally soft argon etching is used to remove the top
10 nm-thick \textit{n}+-doped germanium layer. The whole device is
shown in Fig. \ref{fig01ab}(a). In order to test the influence of
the MgO tunnel barrier (resp. sample annealing), similar devices
without MgO (resp. without annealing) were processed and studied.
Ferromagnetic resonance and inverse spin Hall effect measurements
were performed in a Br\"{u}ker ESP300E $X$-band CW spectrometer with a
cylindrical Br\"{u}ker ER 4118X-MS5 cavity. The measurement geometry is depicted
in Fig. \ref{fig01ab}(b). Complementary measurements of
FMR lines at different frequencies between 2 and 24 GHz were
performed in a stripe-line vector network analyzer system.

\section{Ferromagnetic resonance and electromotive force}

\subsection{Ferromagnetic resonance (FMR)}

\begin{figure}[h!]
\includegraphics[width=6 cm]{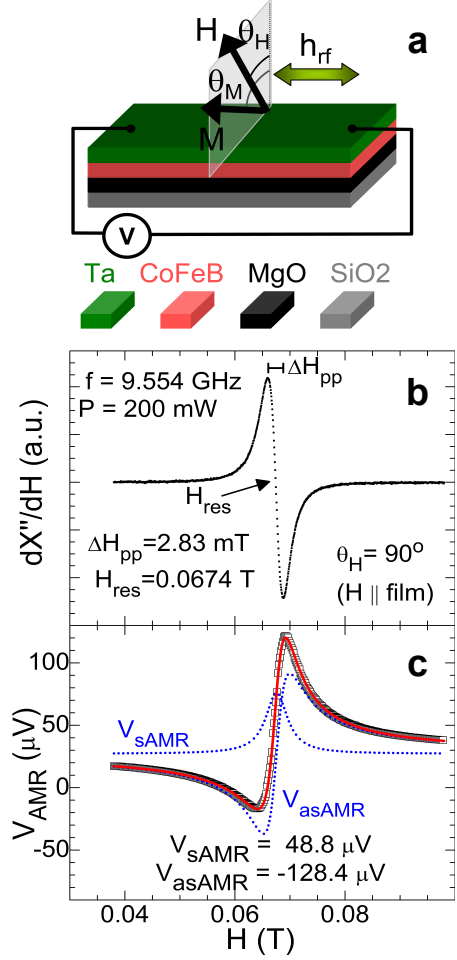}
\caption{(Color online) (a) Schematic drawing of the reference sample
Ta(5nm)/CoFeB(5nm)/MgO(3.3nm)//SiO$_2$. Electrical contacts are directly made on top of the metallic layer. The FMR spectrum (b) and the electromotive force (c) have been measured simultaneously by placing the sample in the EPR cavity. The red line in (c) is
the fit according to Eq. (\ref{eqVamr}). Symmetric
and asymmetric contributions are also shown separately in blue dotted curves.}
\label{fig1}
\end{figure}

Fig. \ref{fig1}(a) shows a schematic drawing of the reference sample
and the definition of the magnetization and external magnetic
field polar angles, $\theta_H$ and $\theta_M$, respectively. The
reference sample is made of a single CoFeB layer grown on SiO$_2$ as
follows: Ta(5nm)/CoFeB(5nm)/MgO(3.3nm)//SiO$_2$. We have inserted a
thin MgO oxide layer between CoFeB and SiO$_2$ in order to make a
comparison with the Ta(5nm)/CoFeB(5nm)/MgO(3.3nm)/Ge
system studied in the next sections.

\begin{figure}
\includegraphics[width=7 cm]{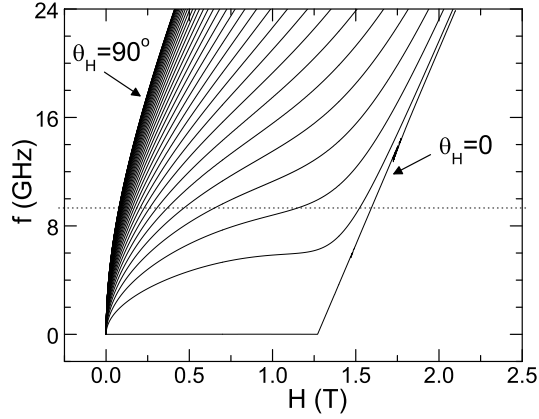}
\caption{Dispersion relationship in FMR experiments according to Eq.
(\ref{eq1}) and (\ref{eqF}) by applying the external magnetic field
at different polar angles $\theta_H$. Curves are shown every
2.5$^{o}$ between \textbf{H} parallel ($\theta_H=90^{o}$) and perpendicular
($\theta_H=0$) to the film plane. The intercept of each curve with
the dotted line ($f\approx 9.55$ GHz) yields the resonance field
$H_{res}$.}\label{fig2dispersion}
\end{figure}

From FMR measurements (Fig.~\ref{fig1}(b)), we can determine the peak-to-peak
linewidth and the resonance field. By sweeping the
external magnetic field \textbf{H} under a microwave excitation of frequency
$f$, the resonance condition is achieved in a ferromagnetic film
when:\cite{Smit1955}

\begin{equation}
\label{eq1} \left( {{\frac{{\omega} }{{\gamma} }}} \right)^{2} =
{\frac{{1}}{{M_{s}^{2}sin^{2}\theta} }}{\left[ {{\left.
{{\frac{{\partial ^{2}F}}{{\partial \theta ^{2}}}}} \right|}_{\theta
_{M0} ,\phi _{M0}}  {\left. {{\frac{{\partial ^{2}F}}{{\partial \phi
^{2}}}}} \right|}_{\theta _{M0} ,\phi _{M0}}  - \left( {{\left.
{{\frac{{\partial ^{2}F}}{{\partial \theta
\partial \phi} }}} \right|}_{\theta _{M0} ,\phi _{M0}} }  \right)^{2}}
\right]}
\end{equation}

\noindent where $\omega = 2\pi f$ is the precession
angular frequency, $\gamma = g\mu _{B} / \hbar $  is the
gyromagnetic ratio with the Land\'e factor $g$ , $\hbar$  is the
reduced Planck constant, $\mu _{B}$ is the Bohr magnetron and
$M_{s}$ is the saturation magnetization of the ferromagnetic film.
The second order partial derivatives of the free energy density are
evaluated at the equilibrium angles $\theta_{M0}$ and $\phi_{M0}$ of
the magnetization \textbf{M} for which: ${\left. {\partial F /
\partial \theta} \right|}_{\theta _{M0} ,\phi _{M0}}  = 0$ and ${\left.
{\partial F / \partial \phi} \right|}_{\theta _{M0} ,\phi _{M0}}  =
0$.  The shape anisotropy in a ferromagnetic polycristalline film ($4\pi M_s$ ) is
usually much larger than in-plane crystalline anisotropy. By recording FMR spectra with the DC magnetic field in the film plane at different azimuthal angles (not shown), we could demonstrate that in-plane anisotropy in the CoFeB electrodes used in this work is indeed negligible with respect to shape anisotropy. Thus we can consider that the free energy density is given
by:

\begin{equation}
\label{eqF} F = - \bf{M}\cdot \bf{H} +2\pi M_{eff}^2 \cos^2 \theta_{M}
\end{equation}

\noindent where the first term is the Zeeman energy and the second
one accounts for shape anisotropy and any other perpendicular uniaxial anisotropy $H_{u\bot}$. The effective saturation magnetization $M_{eff}$ is thus defined as: $4\pi M_{eff}=4\pi M_{s}+H_{u\bot}$. By minimizing numerically $F$ we
can obtain the magnetization equilibrium angles: $\phi_{M0}$ and
$\theta_{M0}$. The resonance field is then given by combining Eq.
(\ref{eq1}) and (\ref{eqF}), it can be plotted as a function of the
external static field orientation ($\theta_H$) and the excitation
frequency as shown in Fig. \ref{fig2dispersion}. $M_{eff}$ and $g$
are extracted from the out-of-plane (OOP) angular dependence of the
resonance field using a least square fit (as shown for instance in
Fig. \ref{fig4oop}(a,d)).

\noindent Analytical solutions of $H_{res}$ can be
obtained for the parallel, $\theta_H=90^{o}$, and perpendicular,
$\theta_H=0$, cases. In the parallel case, we find:

\begin{equation} \label{eqKittel} 
\left({\frac{\omega}{\gamma}}\right)^2 =H_{res}(4\pi M_{eff}+H_{res})
\end{equation}

The frequency dependence of the FMR linewidth also allows calculating
the Gilbert damping constant $\alpha$ using the following
expression:\cite{Heinrich1985}

\begin{equation} \label{eqalpha} \
\Delta H_{pp} =\Delta H_{0}+\Delta H_{G}=\Delta
H_{0}+\frac{2}{\sqrt{3}}\frac{ \omega}{\gamma} \alpha
\end{equation}

\noindent where the peak-to-peak linewidth, $\Delta H_{pp}$,is
measured when $\bf{H}$ is applied parallel to the film plane
($\theta_H=90^o$). The $\Delta H_{0}$ term accounts for the
frequency-independent contributions due to inhomogeneities in the
ferromagnetic layer and $\Delta H_{G}$ is the FMR linewidth due to
the Gilbert damping. As shown experimentally in
section~\ref{sec:Results}, $\Delta H_0<<\Delta H_G$ at high
frequency and we systematically neglect this contribution to the FMR
linewidth. Moreover, the OOP angular dependence of the peak-to-peak
linewidth $\Delta H_{pp}$ at a given frequency can be written:

\begin{equation} \label{eqDHppline}
\Delta H_{pp} = \Delta H_{G}+\Delta H_{\theta}
\end{equation}
where the Gilbert contribution can be calculated from:\cite{Goryunov1995}
$\Delta H_{G}=(2/\sqrt{3})\alpha
(\omega/\gamma)/\cos(\theta_H-\theta_M)$, 
and $\Delta H_{\theta}=|dH_{res}/d\theta_H|\Delta\theta$ is the
angular dispersion of the perpendicular anisotropy and demagnetizing field ($4\pi M_{eff}$) due to inhomogeneities in the FM layer. We show in the
following that $\alpha$ and $\Delta \theta$ can
be extracted from the frequency and OOP angular dependences of
$\Delta H_{pp}$.

\subsection{Electromotive force measured on the reference sample}

The electromotive force generated in the ferromagnetic layer
and shown in Fig. \ref{fig1}(c) is simultaneously recorded with the FMR spectrum.
The origins of this electromotive force are the anisotropic magnetoresistance (AMR)
\cite{Ando2011a,Mosendz2010b,Azevedo2011} and anomalous Hall effect which manifest at the resonance
field. At the resonance field, the precessing magnetization induces a time varying resistivity of the ferromagnetic layer which combines with the radiofrequency induced currents (along $x$ in Fig. \ref{fig01ab}(b)) to produce a DC voltage. The radiofrequency currents are likely produced within the metallic layer by the non-vanishing radiofrequency electric field in the cavity at the sample level. In our set-up geometry, the electromotive force is measured as a voltage along $y$ (see Fig. \ref{fig01ab}(b)) and thus in the transverse Hall geometry. Therefore we measure the planar Hall effect (PHE) and the anomalous Hall effect (AHE) in CoFeB. It was proposed and shown by Azevedo \textit{et al.}
\cite{Azevedo2011} and Harder \textit{et al.} \cite{Harder2011} that the resulting
voltage is well described by both a symmetric and
an asymmetric contributions. In the CoFeB reference film, the asymmetric component is dominant. The electromotive force can be written as (see Appendix A):

\begin{eqnarray} \label{eqVamr}
V &=& V_{offset}+V_{PHE}+V_{AHE}\nonumber \\
&=& V_{offset}+V_{sAMR}\frac{\Delta H^2}{(H-H_{res})^2+\Delta
H^2}+V_{asAMR}\frac{-\Delta H (H-H_{res})}{(H-H_{res})^2+\Delta H^2}
\end{eqnarray}

\noindent where $V_{sAMR}$ (resp. $V_{asAMR}$) is the amplitude of the
symmetric (resp. asymmetric) contribution to the electromotive
force. We have taken into account a non-resonant offset voltage $V_{offset}$,
$H_{res}$ is the resonance field and $\Delta H=(\sqrt{3}/2)\Delta
H_{pp}$ . $V_{sAMR}/V_{asAMR}=-1/\tan\psi$ where
$\psi$ is the phase shift between the radiofrequency current
and the magnetization.\cite{Azevedo2011,Harder2011} The symmetric
and asymmetric contributions as well as the offset voltage are proportional to the microwave
power. It means that  $V_{offset}$, $V_{sAMR}$ , and
$V_{asAMR}$ are proportional to $h_{rf}^2$, where $h_{rf}$ is the
microwave magnetic field strength.

\subsection{Spin pumping and inverse spin Hall effect in germanium}

Here we consider the device shown in Fig. \ref{fig01ab}. Under
radiofrequency excitation the magnetization precession of the
ferromagnetic layer pumps spins to the non-magnetic germanium layer
(N) and the corresponding spin current generates an electric field
in Ge due to ISHE:
$\bf{E_{ISHE}}\propto\bf{J_{S}}\times\bf{\sigma}$. $\bf{J_{S}}$ is
the spin-current density along $z$ and $\bf{\sigma}$ its spin
polarization vector. This electric field $\bf{E_{ISHE}}$ is
converted into a voltage $V_{ISHE}$ between both ends of the Ge
channel.\cite{Ando2011a} In the case of germanium we overcome the
conductivity mismatch issue by inserting a thin MgO tunnel barrier
(I) between Ge and CoFeB. This additional interface resistance
allows for spin accumulation in the germanium conduction band. As a
consequence of spin pumping, the damping constant, $\alpha_{FM/I/N}$,  is enhanced with respect to the
one of the reference sample, $\alpha_{FM/I}$. The real part of the tunnel spin mixing
conductance, $g_{t}^{\uparrow\downarrow}$, is given
by:\cite{Tserkovnyak2005,Mosendz2010b}

\begin{equation} \label{eqgmix}
g_{t}^{\uparrow\downarrow}= \frac{4\pi M_{eff}t_{F}}{g\mu_B}
(\alpha_{FM/I/N}^{}-\alpha_{FM/I}^{})
\end{equation}

where $t_F$ is the CoFeB thickness. When the static magnetic field
is applied parallel to the interface ($\theta_{H}$=90$^{\circ}$),
the spin-current density at the interface between CoFeB/MgO and Ge,
$j_S^0$ is given by:\cite{Ando2011a}

\begin{equation} \label{eqjs0}
j_{S}^{0}=\frac{g_{t}^{\uparrow\downarrow}\gamma^{2}\hbar
h_{rf}^{2}}{8\pi\alpha^{2}}\left[\frac{4\pi M_{eff}\gamma +\sqrt{(4\pi
M_{eff}\gamma)^{2}+4\omega^{2}}}{(4\pi
M_{eff}\gamma)^{2}+4\omega^{2}}\right]
\end{equation}

\noindent where $h_{rf}$ is the strength of the microwave magnetic
field into the resonance cavity. $h_{rf}$ is calculated by measuring
the $Q$ factor of the resonance cavity $Q=f/\Delta f$, where $\Delta
f$ is the width at half maximum of the frequency distribution when
the sample is placed into the cavity. To measure $\Delta f$ we use a
second frequencemeter in series with the first one. The voltage
$V_{ISHE}$ due to the inverse spin Hall effect is
always symmetric with respect to the resonance field and its
amplitude is discussed in Ref. \cite{Ando2011a,Mosendz2010b,Azevedo2011}. We then modify the
equivalent circuit used in Ref. \cite{Ando2011a} and refined the model used in Ref. \cite{Jain2012} to account for
electron transport through the tunnel and Schottky barriers back to
the FM (see Appendix B). Then the ISHE voltage in
our system is given by:

\begin{equation} \label{eqVishe}
V_{ISHE}=\frac{w_F}{t\sigma+t_N\sigma_N}\left[1+\frac{t\sigma}{t_N\sigma_N}\frac{2\lambda}{w_F}\tanh\left(\frac{w_F}{2\lambda}\right)\right]\theta_{SHE}l_{sf}^{cb}\tanh\left(\frac{t_{N}}{2l_{sf}^{cb}}\right)\left(\frac{2e}{\hbar}\right)j_{S}^{0}
\end{equation}

\noindent where $w_F$ is the width of the ferromagnetic electrode
(150 $\mu$m), $t_{N}$ (resp. $t$) is the Ge (resp. Ta/CoFeB)
thickness, $\sigma_N$ (resp. $\sigma$) is the Ge (resp. Ta/CoFeB)
conductivity. $t\sigma=t_F\sigma_F+t_{Ta}\sigma_{Ta}$ where $t_F$ and $\sigma_F$ (resp. $t_{Ta}$ and $\sigma_{Ta}$) are the thickness and conductivity of the CoFeB (resp. Ta) layer. $\lambda$ depends on the resistance-area product $RA$ of the interface between CoFeB/MgO and Ge as:

\begin{equation} \label{eqdelta}
\left(\frac{1}{\lambda}\right)^{2}=\left(\frac{1}{t\sigma}+\frac{1}{t_N\sigma_N}\right)\frac{1}{RA}
\end{equation}

In order to estimate the $V_{ISHE}$ magnitude, the electromotive force and the
ferromagnetic spectrum are measured simultaneously. The measured voltage might have one
symmetric, $V_s$, one asymmetric, $V_{asAMR}$, and one offset
contributions. The raw data will be fitted with:

\begin{equation} \label{eqV2}
V = V_{offset}+V_{s}\frac{\Delta H^2}{(H-H_{res})^2+\Delta
H^2}+V_{asAMR}\frac{-\Delta H (H-H_{res})}{(H-H_{res})^2+\Delta H^2}
\end{equation}

Note that Eq. (\ref{eqV2}) is similar to Eq. (\ref{eqVamr}) but in the
presence of spin pumping the symmetric voltage is:
$V_s=V_{ISHE}+V_{sAMR}$.

\section{Results}
\label{sec:Results}

\subsection{Reference sample}

\begin{figure}[h!]
\includegraphics[width=12 cm]{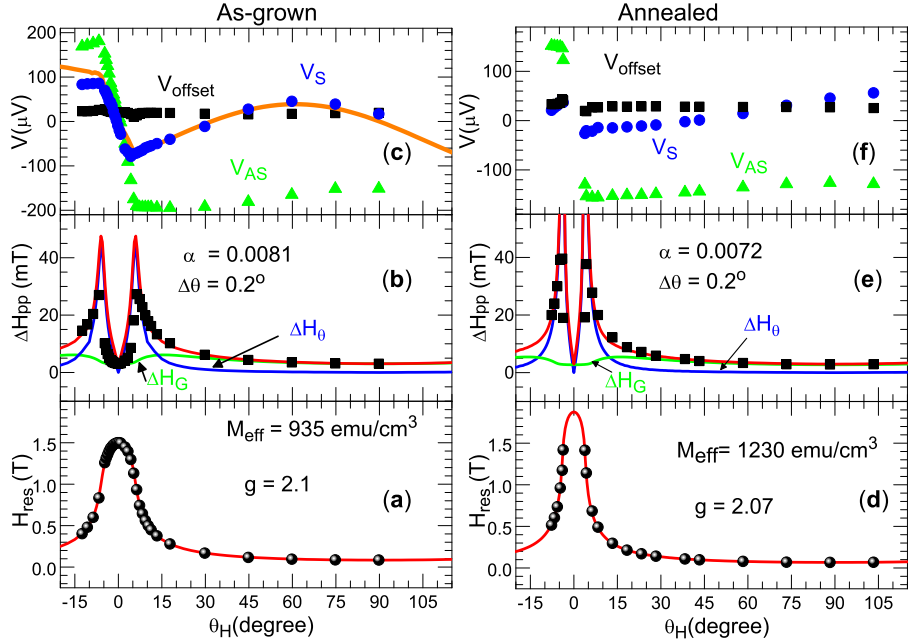}
\caption{(Color online) Angular out-of-plane dependence of $H_{res}$
(a,d), $\Delta H_{pp}$ (b,e), and amplitude of the electromotive force
(c,f) according to Eq. (\ref{eqVamr}). Samples are multilayers-see
Fig. \ref{fig1}(a)- of Ta(5nm)/CoFeB(5nm)/MgO(3.3nm)//SiO2 as-grown
(a-c) and annealed (d-f). Red curves in (a,d) are numerical fits as
explained above. The curves in (b,e) are numerical calculations to the
$\Delta H_{pp}$ contributions according to Eq. (\ref{eqDHppline}).
The solid curve in (c) is a fit using the
$V_{PHE}(\theta_H)$ formula of appendix A, Eq.
(\ref{eqVPHE}). }\label{fig4oop}
\end{figure}

Fig. \ref{fig4oop} shows the OOP dependence of the resonance
field, peak-to-peak linewidth and electromotive force on the
as-grown and annealed Ta(5nm)/CoFeB(5nm)/MgO(3.3nm)//SiO$_2$
reference samples. From the angular dependence of $H_{res}$, we
obtain the effective saturation magnetization ($M_{eff}$) and the $g$ factor and
from $\Delta H_{pp}$ we obtain the damping constant ($\alpha$) and
the angular dispersion ($\Delta\theta$). The angular dependence of
the peak-to-peak linewidth can be calculated using the following
method: after fitting numerically the OOP dependence of the
resonance field (Fig. \ref{fig4oop}(a,d)) we use $M_{eff}$ and $g$ to
calculate the theoretical dispersion relationship between $f$ and
the external magnetic field for different $\theta_H$ angles. This is
shown in Fig. \ref{fig2dispersion} where the dotted line
corresponds to the frequency at which the measurements are
performed. The intercept of each curve with the dotted line gives
the value of the resonance field $H_{res}$ along with the
equilibrium polar angle $\theta_{M0}$ of $\bf{M}$ at different
$\theta_H$ values. The OOP linewidth angular dependence is shown in
Fig. \ref{fig4oop}(e) and fitted using Eq. (\ref{eqDHppline}).
In addition, in Fig. \ref{fig4oop}(c), we have used the $V_{PHE}(\theta_H)$ formula of Appendix A to fit the OOP angular dependence of the symmetric voltage contribution to the electromotive force in the as-grown reference sample. The OOP angular dependence of the symmetric voltage in
the Ge-based device ($V_{ISHE}$) clearly shows a different behavior (see Fig.
\ref{fig8Geoop}(c) and \ref{fig8Geoop}(f)).


\begin{figure}[h!]
\includegraphics[width=6 cm]{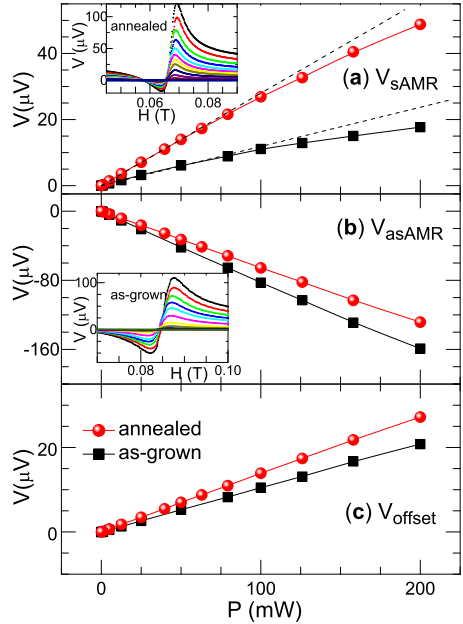}
\caption{(Color online) Power dependence of the symmetric ($V_{sAMR}$) and asymmetric ($V_{asAMR}$) contributions
to the electromotive force according to eq. (\ref{eqVamr}). Samples are
multilayers of Ta(5nm)/CoFeB(5nm)/MgO(1.1nm)//SiO2 as grown
(squares) and annealed (circles). Solid lines are guides for the
eyes and dashed lines in (a) show the non linear behavior of
$V_{sAMR}$ in all the experimental frequency range. The insets show the
electromotive force measured under different power excitations.}\label{fig5pow}
\end{figure}

We have also recorded the power dependence of both the FMR signal
and the electromotive force when $\bf{H}$ is applied parallel to the
film plane. The results are shown in Fig. \ref{fig5pow}. The
electromotive force was fitted according to Eq. (\ref{eqVamr}). $V_{offset}$ and $V_{asAMR}$ depend linearly on the
applied power in the whole power range whereas $V_{sAMR}$ slightly
deviates from the linear behavior for high powers. Nevertheless we
note that $V_{sAMR}<<V_{asAMR}$ in both reference samples. The
sample sizes are $\sim2\times3.5$ mm$^2$ for the as-grown sample and
$\sim2\times1.5$ mm$^2$ for the annealed one.
For the same RF power of 200 mW and the field applied parallel to the film plane ($\theta_H$=90$^{\circ}$), we found $V_{asAMR}\approx$-159 $\mu V$ for the as-grown sample and $V_{asAMR}\approx$-64.2 $\mu V$ for the annealed one. Since $V_{asAMR}$ depends linearly on the ferromagnetic electrode width, we can estimate the expected $V_{asAMR}$ value for the CoFeB bar of Fig.~\ref{fig01ab}(a): $V_{asAMR}\approx$6.8 $\mu V$ for the as-grown sample and $\approx$6.4 $\mu V$ for the annealed one. The expected symmetric contribution to the electromotive force $V_{sAMR}$ will then be almost one order of magnitude less. Furthermore, in the device of Fig.
\ref{fig01ab}, the electrical contacts are no more made on the
metallic multilayer, as show in Fig. \ref{fig1}(a), but on Au/Ti ohmic
contacts on top of Ge which would reduce the PHE and AHE contributions.

\begin{figure}[h!]
\includegraphics[width=6 cm]{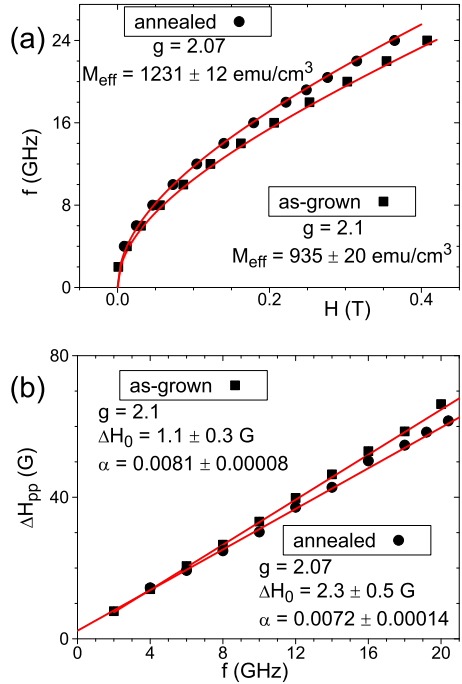}
\caption{(Color online) FMR dispersion relationship (a) and
frequency dependence of the peak-to-peak linewidth (b) for the parallel
case. Black dots are for the annealed sample and squares for the as-grown one. Solid red lines are fits
according to Eq. (\ref{eqKittel}) in (a), and Eq.(\ref{eqalpha}) in
(b).}\label{figDampingRefer}
\end{figure}

Fig. \ref{figDampingRefer} shows the frequency dependence of the
resonance field (a) and peak-to-peak linewidth (b) of the as-grown
and annealed reference samples. In both figures, we have used the
$g$ factors deduced from the OOP angular dependence of $H_{res}$ (see
Fig. \ref{fig4oop}(a) and \ref{fig4oop}(d)) and adjusted the $M_{eff}$ and $\alpha$
values according to eq. (\ref{eqKittel}) and (\ref{eqalpha})
respectively to fit the curves. The frequency independent part of
the peak-to-peak linewidth $\Delta H_{0}$ which is due to
inhomogeneities in the magnetic layer is very weak in both samples.
We find 1.1 Oe for the as-grown sample and 2.3 Oe for the annealed
one which confirms that $\Delta H_{0}<<\Delta H_{G}$ at a frequency
close to 9 GHz. Moreover the effective perpendicular anisotropy (4$\pi M_{eff}$) increases
from 1.175 T up to 1.545 T upon annealing as recently
reported.\cite{Wang2011} Both samples exhibit very low damping
constants comparable to the ones found in Ref. \cite{Liu2011JAP}.
Interestingly the peak-to-peak linewidth and damping constants
decrease upon annealing in contrast with other
results.\cite{Chen2012} It means that we have effectively reduced
the intrinsic inhomogeneities of the CoFeB electrode by annealing.
In particular, the annealing process did not promote chemical
inter-diffusion at the interfaces with CoFeB as found in thinner
CoFeB films in magnetic tunnel junctions.\cite{Yang2012}

\subsection{Spin pumping at the ferromagnet/Germanium interface}

\subsubsection{CoFeB/Ge Interface}

\begin{figure}[h!]
\includegraphics[width=12 cm]{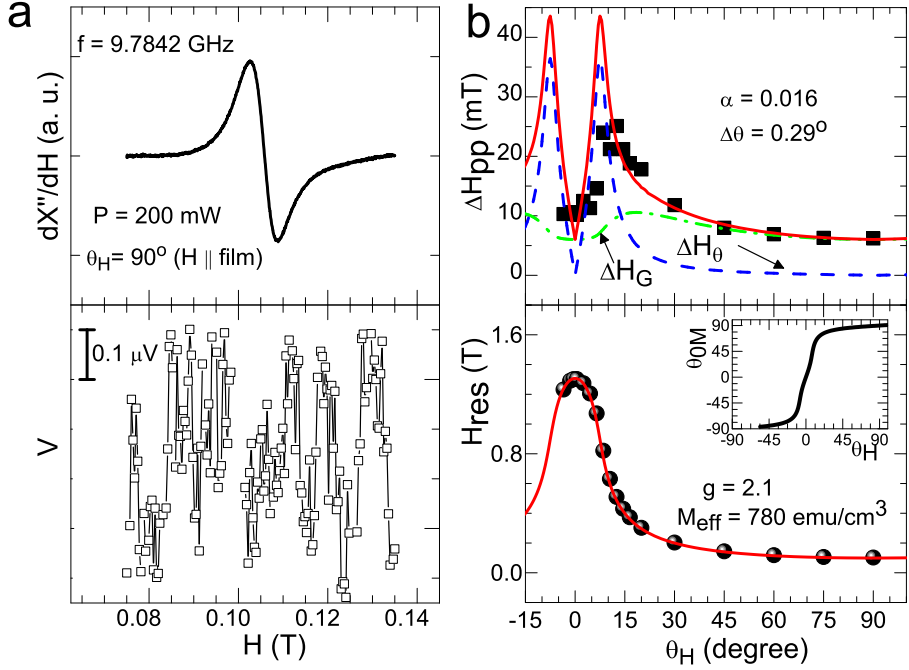}
\caption{(Color online) Results on the device without MgO tunnel barrier
and without annealing: (a) FMR line along with the voltage measured
in the parallel case. (b) OOP angular dependence of the resonance
field and peak-to-peak linewidth with their numerical fits. The inset
shows the equilibrium angle of the magnetization as a function of $\theta_H$.
Similar devices without MgO barrier and after annealing process do not
exhibit ferromagnetic resonance.}\label{figGe_CFB}
\end{figure}

\begin{figure}[h!]
\includegraphics[width=6 cm]{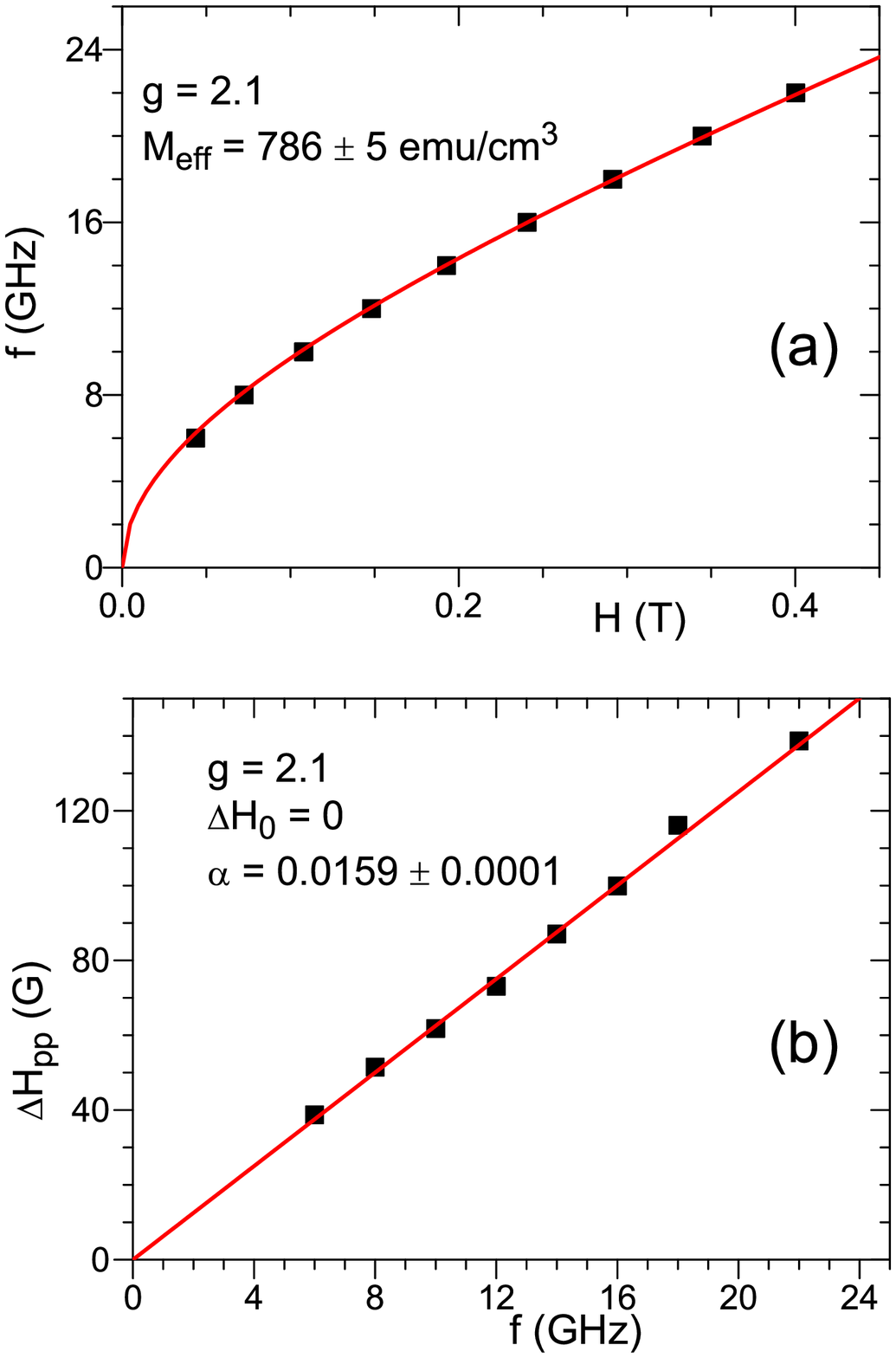}
\caption{(Color online) FMR dispersion relationship (a) and
frequency dependence of the peak-to-peak linewidth (b) in the parallel
case. The sample is a Ta(5nm)/CoFeB(5nm)/Ge device
without MgO tunnel barrier. The red solid lines are fits according to Eq.
(\ref{eqKittel}) in (a), and Eq.(\ref{eqalpha}) in (b)
}\label{figDampingCFBGe}
\end{figure}

In this section, we consider the device shown in Fig.~\ref{fig01ab}(a) where the CoFeB electrode
 has been directly grown on the Ge film without tunnel barrier. The FMR line and the corresponding
 electromotive force are shown in Fig.~\ref{figGe_CFB}(a). A clear absorption is observed in
 the FMR spectrum whereas the electromotive force at the resonance field is negligible. Hence, in the measuring geometry of Fig.~\ref{fig01ab}(a) where the voltage is directly probed on the germanium layer, we do not detect the PHE and the AHE in the CoFeB ferromagnetic layer at the resonance. The angular dependence of the resonance field and peak-to-peak linewidth are displayed
 in Fig.~\ref{figGe_CFB}(b). The frequency dependence of $\Delta H_{pp}$ and $H_{res}$ are shown in Fig. \ref{figDampingCFBGe}.
 First, the effective CoFeB saturation magnetization $M_{eff}$ is lower than in the
 reference sample: this is probably due to the intermixing between CoFeB and Ge at the
 interface. In the same way, the larger damping constant $\alpha$ may be due to interface inhomogeneities
 as a consequence of intermixing and not to spin pumping since no electromotive force is
 observed.
 We have then performed the same measurements on the annealed sample.
 In that case, both the ferromagnetic resonance signal and the electromotive force
 vanish and the CoFeB film has completely diffused into the Ge layer. These
 results show that the MgO tunnel barrier is not only necessary to overcome
 the conductivity mismatch issue but also to prevent the intermixing between CoFeB and Ge at the interface.

\subsubsection{CoFeB/MgO/Ge Interface}

\begin{figure}[h!]
\includegraphics[width=12 cm]{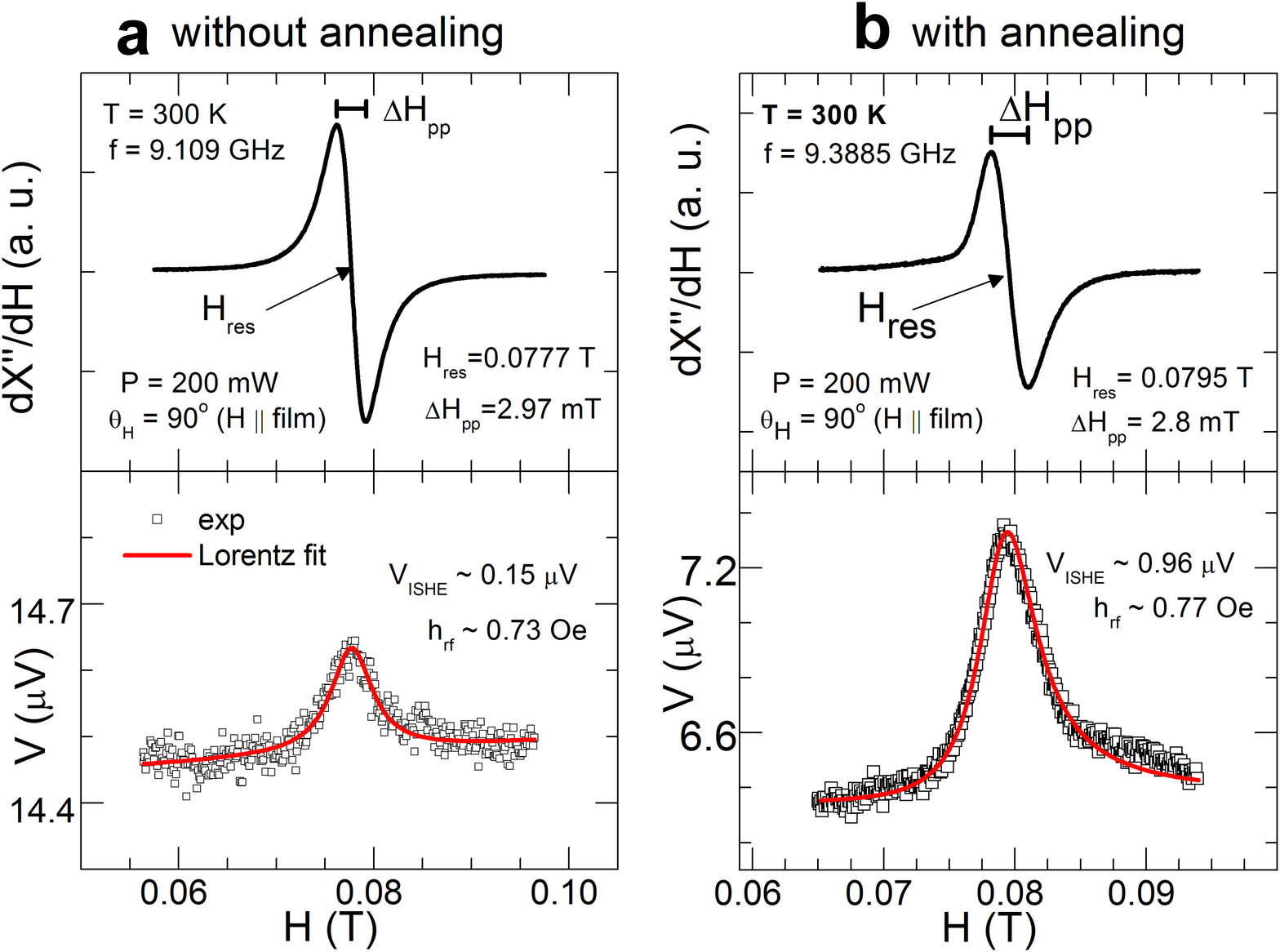}
\caption{(Color online) FMR spectrum and electromotive force measured simultaneously on the as-grown CoFeB/MgO/Ge sample (a) and the annealed one (b). The magnetic field is
applied parallel to the CoFeB bar. There is clearly a voltage peak at
the resonance condition and an enhancement of that peak by annealing.}\label{figGe_MgO_CFB}
\end{figure}

\begin{figure}[h!]
\includegraphics[width=12 cm]{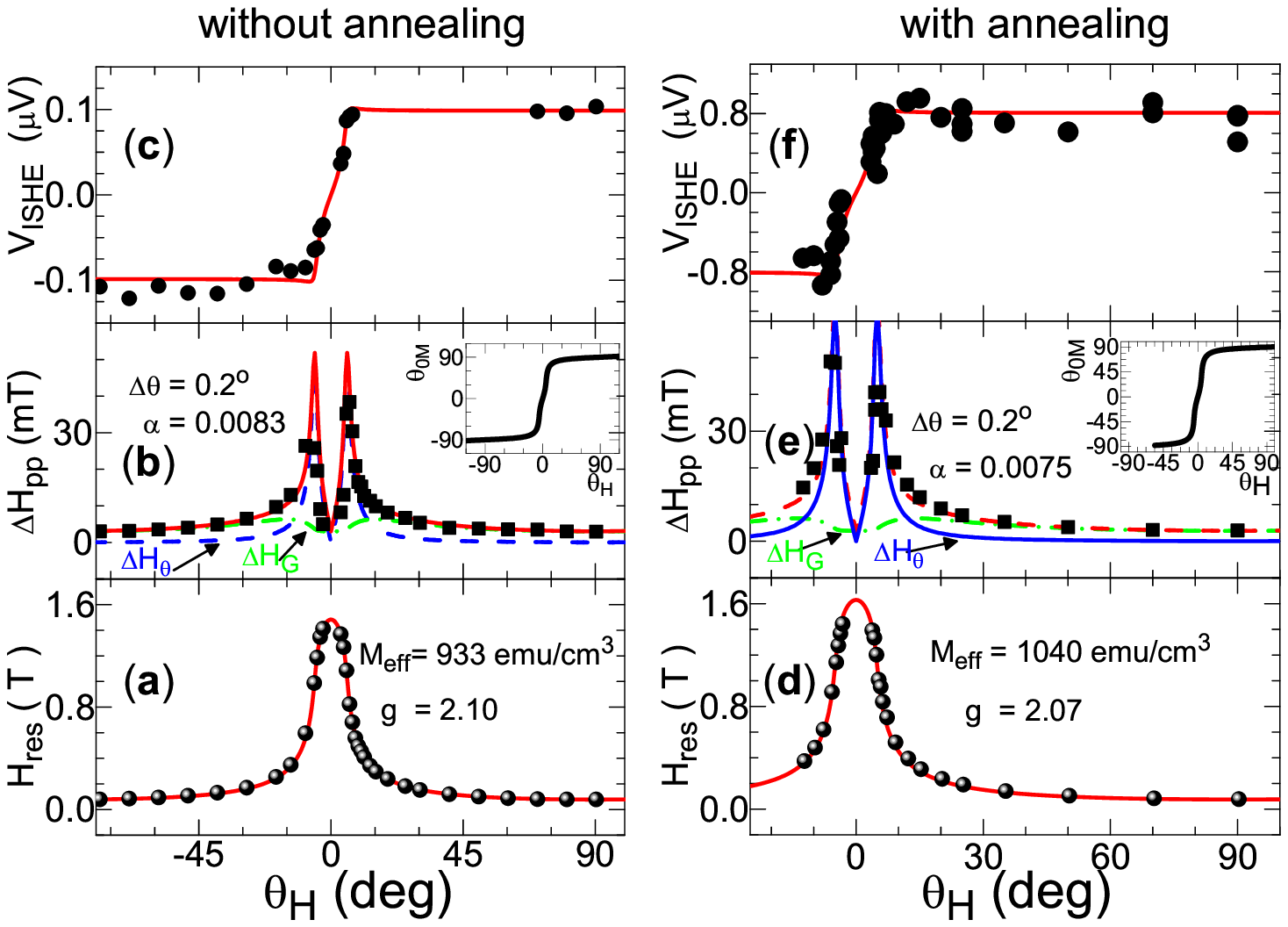}
\caption{(Color online) Angular out-of-plane dependence of the
resonance field (a,d), peak-to-peak linewidth (b,e), and ISHE
voltage (c,f) on the as-grown CoFeB/MgO/Ge sample (a-c) and the annealed one (d-f). The numerical fits are in solid lines.
The insets in (b,e) show the equilibrium angle of the magnetization as a function of
$\theta_H$.}\label{fig8Geoop}
\end{figure}

We now consider the same device as in the previous section but with a thin
MgO tunnel barrier inserted between CoFeB and Ge as shown in
Fig.~\ref{fig01ab}(a). The FMR spectrum and the corresponding
electromotive force are shown in Fig.~\ref{figGe_MgO_CFB}(a) for the
as-grown sample and Fig.~\ref{figGe_MgO_CFB}(b) for the annealed one.
Here a clear electromotive force is detected at the resonance field
in both cases. The red line is the fit according to Eq. (\ref{eqV2})
considering a single symmetric contribution. Moreover by annealing
we observe an enhancement of the electromotive force signal. In
Fig.~\ref{fig8Geoop}, the OOP angular dependence of the resonance
field, peak-to-peak linewidth and the amplitude of the electromotive
force of both samples are displayed. Like in the previous section, the complete
analysis of these data yields $M_{eff}$, $g$, $\alpha$, and $\Delta
H_0$. The solid lines in Fig.~\ref{fig8Geoop}(c) and Fig.~\ref{fig8Geoop}(f) are fits
according to the formula:\cite{Ando2011b} $V_{ISHE}(\theta_H)\propto
sin(\theta_{M0})\{[(H_{res}/(4\pi
M_{eff}))cos(\theta_{M0}-\theta_H)-cos(2\theta_{M0})]/[(2H_{res}/(4\pi
M_{eff}))cos(\theta_{M0}-\theta_H)-cos(2\theta_{M0})-cos^2(\theta_{M0})]^2\}$.

We also measured the FMR spectrum in the parallel case at different
frequencies on both devices (not shown). The frequency dependence of
$\Delta H_{pp}$ always shows a linear behavior with a very low
$\Delta H_0$ value showing that the Gilbert-type effect is the
dominating contribution to the damping in all the samples studied.

In Table \ref{tableAlpha}, we can clearly see that the annealing process increases the perpendicular
magnetic anisotropy of the system (enhancement of $M_{eff}$) and
reduces the intrinsic damping constant. Spin pumping in Ge leads to an increase of the damping constant ($\alpha_{CoFeB/MgO/Ge}$)
with respect to that of the reference system ($\alpha_{CoFeB/MgO}$).

\begin{table}[htbp]
\begin{center}
\begin{tabular}  {lccccc}
\hline
     & \multicolumn{2}{c}{$\alpha$ ($10^{-3}$)} &\multicolumn{2} {c}{~~~~~~~~~~~~~~$M_{eff}
    (emu/cm^3)$}\\  \cline{2-6}
    & CoFeB/MgO ref. & ~~n-Ge device &  ~~~    & CoFeB/MgO ref. & ~~n-Ge device  \\
   as-grown~ & $8.1\pm 0.08$ & $8.3\pm0.06$& & $935\pm20$ & $940\pm15$  \\
   annealed~ & $7.2\pm0.14$ & $7.5\pm0.27$ & & $1230\pm12$ & $1040\pm20$
   \\ \hline
\end{tabular}
  \caption{Damping constant and effective saturation magnetization of the CoFeB/MgO/Ge system and CoFeB/MgO reference sample.} \label{tableAlpha}
\end{center}
\end{table}
%

\begin{figure}[h!]
\includegraphics[width=6 cm]{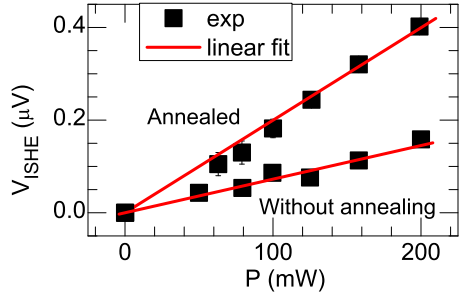}
\caption{(Color online) Power dependence of the
ISHE voltage measured on the Ta/CoFeB/MgO/Ge device with and without
annealing. The voltage depends linearly on the excitation power.}\label{fig9PowdepGe}
\end{figure}

The power dependence of the $V_{ISHE}$ amplitude when the external
DC magnetic field is applied parallel to the FM layer is shown in
Fig.~\ref{fig9PowdepGe} where the solid line is a linear fit. Such
linear behavior accounts well for the $h^{2}_{rf}$ dependence of the
V$_{ISHE}$ since the microwave power is proportional to the square of the
rf magnetic field ($P\propto h^{2}_{rf}$).

All these results support the fact that the measured electromotive
force is due to spin pumping from the CoFeB electrode and inverse
spin Hall effect in germanium.

\subsubsection{Estimation of the spin Hall angle in n-Ge at room temperature}

In order to estimate the spin Hall angle $\theta_{SHE}$ in $n$-Ge,
we have calculated the tunnel spin mixing conductance according to Eq.
(\ref{eqgmix}). We found: $g_{t}^{\uparrow\downarrow}=
6.1\times 10^{17}$ $m^{-2}$ for the device with the as-grown CoFeB
layer and $1.0\times 10^{18}$ $m^{-2}$ for the device with the
annealed CoFeB layer. The spin-current density at the interface $j_{S}^{0}$, when
$\theta_{H}$=90$^{\circ}$, is calculated using Eq. (\ref{eqjs0}) where
the CoFeB effective saturation magnetization $M_{eff}$, the gyromagnetic ratio $\gamma$,
and the damping factor $\alpha$ were deduced from FMR measurements. The results are reported in Table \ref{tableAll}.
For a power of 200 mW, the microwave magnetic field ($h_{rf}$) was
measured with the sample inside the resonator cavity.
The $V_{ISHE}$ amplitude is calculated according to Eq.
(\ref{eqVishe}) with  the width of the ferromagnetic electrode
$w_F$=150 $\mu$m, the FM thickness $t_F=5$ nm, the Ge channel
thickness $t_{Ge}$=40 nm and the conductivities
and the resistance-area product of the interfase $RA_{CoFeB/MgO/Ge}$
given in Table \ref{tableAll}.
The spin diffusion length in the semiconductor channel is
$l_{sf}^{Ge}\approx$1.3 $\mu$m (Ref. \cite{Jain2012}). The conductivities
(including the interface $RA$ value) and the spin diffusion length were measured independently
on the same device (Ref. \cite{Jain2012}).

\begin{table}[htbp]
\begin{center}
\begin{tabular}  {lcc}
\hline
     & \multicolumn{1}{c}{Not annealed} &\multicolumn{1} {c}{annealed}\\  \cline{2-3}
     $g_{t}^{\uparrow\downarrow}$ $(10^{18}$ $m^{-2}$)& 0.6 & 1.0  \\
   $h_{rf}$ (G) & 0.73 & 0.77   \\
   $j_{s0}$ ($nJ/m^{2}$)& 0.047 & 0.11 \\
    $\sigma_{F} (\Omega-cm)^{-1}$ & 9440 & 1250  \\
    $\sigma_{N} (\Omega-cm)^{-1}$ & 270 & 270 \\
    $RA (10^{-1} \Omega cm^{2})$ & 1.047 & 1.085 \\
    $''\theta_{SHE}~''$ & 0.0011 & 0.00044 \\
   \hline
\end{tabular}
  \caption{Measured and calculated parameters on the CoFeB/MgO/Ge sample from
   spin pumping and inverse spin Hall effect measurements in order to estimate $\theta_{SHE}$ of $n-Ge$ according to Eq. (\ref{eqVishe}).
    The conductivities and $RA$ products
    were measured separately.} \label{tableAll}
\end{center}
\end{table}

We then estimate the spin Hall angle in $n$-Ge from the annealed
sample at room temperature: $\theta_{SHE}\approx 0.0011$, which is
of the same order of magnitude as in $n$-GaAs (0.007 in Ref.
\cite{Ando2011b}) and one order of magnitude larger than in $p$-Si
(0.0001 in Ref. \cite{Ando2012}). In a similar way we could estimate
the spin Hall angle in $n$-Ge using the data from the as-grown
sample and found: $\theta_{SHE}\approx$0.00044. Such a difference might come either from the error bars
and/or from the phenomenological model we have used here. We  have
measured several annealed devices from the same batch and found
$\theta_{SHE}$ between 0.0010 and 0.0012 which gives an estimation
of the error bar. We thus conclude that the phenomenological model
we use to estimate $\theta_{SHE}$ is not adapted to our system. In
particular, this model does not account for the presence of
interface states between MgO and Ge. We have shown in a previous
work that interface states play a crucial role in the spin injection
mechanism.\cite{Jain2012} Electrical spin injection into Ge proceeds
by two-step tunneling: the electrons tunnel from the FM to the
localized interface states (IS) through the MgO barrier and from the
IS to the Ge conduction band through the Schottky barrier. Because
spin flips occur into interface states, the spin accumulation (hence
the spin current) is drastically reduced in the Ge conduction band.
By annealing, the density of interface states is reduced and direct
spin injection into the Ge conduction band is favored. As a
consequence, the spin current in the as-grown sample $j_{s}^{0}$ is
reduced as compared to the spin current in the annealed sample which
leads to the underestimation of $\theta_{SHE}$ as found
experimentally. We thus give in the next section a microscopic model
accounting for the presence of the tunnel barrier and interface
states to accurately describe spin pumping and ISHE in germanium.

\section{Discussion}

\begin{figure}[h!]
\includegraphics[width=10 cm]{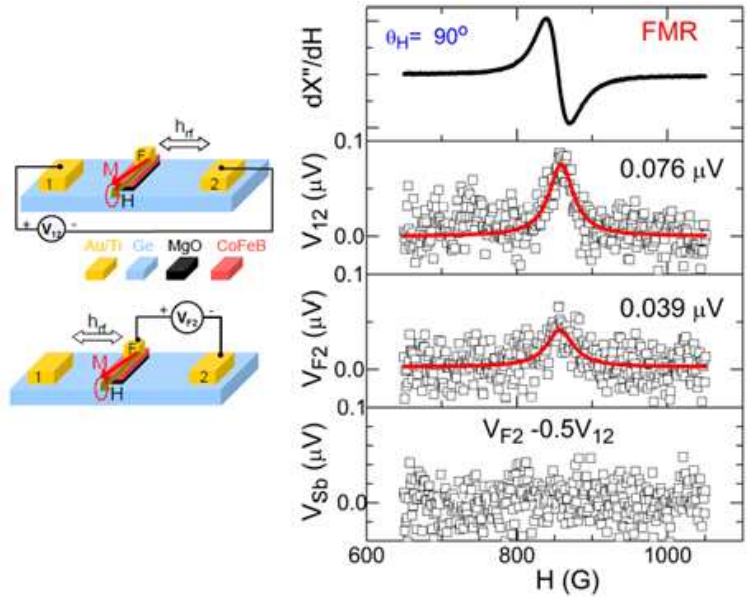}
\caption{(Color online) FMR espectrum along with the
transverse voltage measured to study ISHE ($V_{12}$) and the
voltage between the FM layer and one of the ohmic contacts ($V_{F2}$).
On the left are shown the contacts geometries. Finally, the expected tunneling spin Seebeck voltage is shown at the bottom right of the figure.} \label{figSeebeck}
\end{figure}

In the as-grown and annealed CoFeB/MgO/Ge samples, we could clearly measure an electromotive force due to ISHE at the ferromagnetic resonance of CoFeB. This photovoltage has a symmetric Lorentzian shape. Furthermore we have shown that all our findings
are in good agreement with the observation of ISHE: symmetrical
behaviour of $V_{ISHE}$ around the resonance field $H_{res}$,
$V_{ISHE}$=0 when the external magnetic field is applied
perpendicular to the film ($\theta_{H}$=0) , $V_{ISHE}$ changes its
sign when crossing $\theta_{H}$=0 (Fig. \ref{fig8Geoop}), and
finally the linear dependence of its amplitude with the microwave
power excitation (Fig. \ref{fig9PowdepGe}). This result clearly
demonstrates the presence of both spin accumulation and related spin
current in the Ge conduction band at room temperature. It was also supported by temperature dependent measurements in a previous work.\cite{Jain2012}
In order to confirm that the photovoltage we measure is really due to ISHE and rule out any spurious effects, we carried out complementary measurements. First we studied the photovoltage in millimeter-sized reference samples (both as-grown and annealed) made of CoFeB/MgO/SiO$_2$ with the voltage probes directly connected to the CoFeB film. In that case, we found a dominant asymmetric voltage contribution with respect to the resonance field. It corresponds to the planar Hall effect in the ferromagnet as a combination of anisotropic magnetoresistance and the rf current induced in the ferromagnet by the non-vanishing electric field from the cavity (see Appendix A). This asymmetric voltage contribution due to PHE could not be detected on the device of Fig.~\ref{fig01ab} with Ge. Moreover the out-of-plane angular dependence of this weak symmetric voltage on the reference sample is different from that of the symmetric voltage we detected in the device of Fig.~\ref{fig01ab} with Ge. To summarize this study on the reference sample, we can claim that the symmetric photovoltage observed in CoFeB/MgO/Ge samples is due to ISHE and not to PHE in the ferromagnet. We also carried out the same measurements on CoFeB/Ge samples to study the effect of the MgO tunnel barrier. Without MgO tunnel barrier, we never detected a photovoltage in Ge. It first proves that the photovoltage due to the PHE in CoFeB is undetectable in Ge. It also shows that the MgO tunnel barrier is necessary to perform spin injection in Ge by spin pumping. Furthermore, as shown in Fig.~\ref{figSeebeck}, we have recorded several voltages on the same CoFeB/MgO/Ge device at the ferromagnetic resonance in order to estimate the tunneling spin Seebeck effect.\cite{LeBreton2012,Jain2012b} Indeed at the ferromagnetic resonance, the CoFeB electrode absorbs part of the incident microwave power which increases its temperature. As a consequence, a vertical temperature difference may appear between the CoFeB electrode and the Ge layer. This temperature gradient may create a tunneling Seebeck voltage and a tunneling spin Seebeck voltage which is only a few percents of the Seebeck voltage.\cite{Jeon2012} The resulting spin current gives rise to ISHE in germanium just like spin pumping. In order to discriminate between spin pumping and the tunneling spin Seebeck effect, we measured the following voltages at the FMR: $V_{12}$, $V_{1F}$ and $V_{F2}$ shown in Fig. \ref{figSeebeck}. The sum of the tunneling Seebeck and tunneling spin Seebeck voltages ($V_{Sb}$) is given by: $V_{Sb}=V_{F2}-\frac{1}{2}V_{12}$ or $V_{Sb}=V_{1F}-\frac{1}{2}V_{12}$. As shown in Fig. \ref{figSeebeck}, $V_{Sb}$ is negligible (below the noise level) which rules out the presence of tunneling spin Seebeck at the ferromagnetic resonance in our system. Therefore spin injection in Ge proceeds by spin pumping and not by tunneling spin Seebeck effect.

We now address the important issue of the microscopic origin of spin pumping effects in Ge through a MgO tunnel barrier from a theoretical point of view. As demonstrated below, the origin of spin pumping into SCs through a tunnel barrier lies in the \textit{evanescent} but however \textit{non-zero} exchange coupling between a band of localized states (LS) and the ferromagnet through the tunnel barrier, nonetheless sufficiently transparent. Indeed, spin-pumping~\cite{tserkovnyak2002,Tserkovnyak2005} in metallic tunnel junctions is expected to fall-off in the absence of any exchange field experienced from the ferromagnet (FM) by the delocalized carriers injected in the non-magnetic metal (N). On the other hand, spin injection into a SC by electrical means, as well as by spin-pumping, requires a tunnel barrier at the interface between both types of materials~\cite{rashba2000,Fert2001} in order to overcome the impedance mismatch issue~\cite{schmidt2000} describing a total diffusive spin current backflow towards the FM. As shown in our experiments, spin-pumping in a semiconductor with a tunnel barrier can be recovered with some conditions. First, the carriers injected by tunneling from a FM contact have to remain localized at the interface between the tunnel barrier and the semiconductor in the timescale of a single magnetization precession. Second, the effective \textit{tunnel exchange} field experienced by the carriers, that we call hereafter $\widetilde{J}$, has to be large enough for the spin to rotate in a timescale of a magnetization precession. These two necessary conditions may be fulfilled within a two-step tunneling picture of spin injection into evanescent (or localized states)~\cite{Tran2009} and in the limit of an effective exchange field larger than a certain lower bound. This will be demonstrated below. The third condition to observe spin-pumping in FM/tunnel barrier/SC systems is a minimum value for the conductance of the Schottky barrier delimiting the two regions \textit{i.e} the evanescent states and the SC channel. A thermal activation may be needed to fulfill this third condition. We thus give an analytical expression for the source term taking into account a two-step tunneling process.

Let us consider the standard theory of spin-pumping at the FM/N interface. The source term is known to be equal to:\cite{tserkovnyak2002,Tserkovnyak2005}

\begin{equation}
{I}_s^p=\frac{\hbar}{4\pi}\left(\mathcal{R}e~g^{\uparrow \downarrow}{\bf{m}}\times \frac{d {\bf{m}}}{dt}+\mathcal{I}m~g^{\uparrow \downarrow}
\frac{d {\bf{m}}}{dt}\right)
\end{equation}

\noindent in the case of a FM/N ohmic contact where ${\bf{m}}$ is the unit magnetization vector and $g^{\uparrow \downarrow}$ the complex spin mixing conductance. The spins pumped into N then create a \textit{diffusive} spin current backflow to the FM according to the three-dimensional spin-dependent transmission matching at the FM/N interface:\cite{brataas2006}

\begin{eqnarray}
I_s^b=\frac{g}{8\pi}\left[2p(\mu_0^F-\mu_0^N)+\mu_s^F -{\bf{m}}.\bf{\mu_s^N}\right]~{\bf{m}}\nonumber \\
-\frac{\mathcal{R}e~g^{\uparrow \downarrow}}{4\pi}{\bf{m}}\times (\bf{\mu_s^N}\times {\bf{m}})+\frac{\mathcal{I}m~g^{\uparrow \downarrow}}{4 \pi}({\bf{m}}\times \bf{\mu_s^N})
\end{eqnarray}

\noindent where $\mu_0^N$, $\bf{\mu_s^N}$ in N and $\mu_0^F$, $\mu_s^F\bf{m}$ in FM are respectively the charge and spin accumulations at the interface. $g$ is the sum of spin-up and spin-down conductances and $p$ is the interfacial spin asymmetry coefficient. This backflow of spin current results in a down-renormalization of the spin current pumped in the non magnetic material as shown in a recent couple of papers.\cite{bauer2012,vanwees2012} The exact form of the corresponding \textit{down renormalization} has to be considered case by case.

The new source term describing spin-pumping in a broad band of evanescent states at the tunnel barrier/SC interface has to involve a small but however non-zero exchange interaction $\widetilde{J}_{exc}$ between localized states and the magnetization $\bf{M}$ (of unit vector $\bf{m}$) of the FM through the tunnel barrier; this exchange interaction couples evanescent wavefunctions inside the barrier. In the following, we will define $\widetilde{J}_{exc}$ in the form: $\widetilde{J}_{exc}=J_0~\exp{-2\kappa d}\approx J_0~ T$ where $J_0$ is the bare \textit{on-site} exchange interaction of the order of the exchange interaction in FM or even larger (about 1~eV), $\kappa$ is the imaginary electronic wavevector in the barrier and $d$ the barrier thickness. $T$ is the tunnel transmission coefficient. We note $\Gamma=\hbar/\tau_n$ the mean energy broadening of the localized states due to the finite carrier lifetime ($\tau_n$) through their escape towards the ferromagnetic reservoir FM. This energy broadening can be expressed \textit{vs.} the localization energy $\epsilon_n$ within the centers and $T$ according to $\Gamma \approx \epsilon_n T$ (Ref. \cite{larkin1987}). Note that the escape towards the semiconductor channel, moderately doped, is generally prohibited in an energy band located downward the Fermi energy.

If one defines three different components of the carrier spin vector $\textbf{s}$ injected in the evanescent states by $s_z$, $s_+=s_xm_x+s_ym_y$ and $s_-=s_ym_x-s_xm_y$, the equation of motion for the injected spin, along the $x$ direction at time $t=0$, in a localized state in the exchange field of the magnetization ${\bf{m}}$ rotating in the ($x$,$y$) plane follows the Heinsenberg evolution for the spin-operator $(i\hbar)d{\bf{s}}/dt=[{\bf{s}},\widetilde{J}_{exc}{\bf{s}}.{\bf{m}}]_{-}$; thus giving \textit{in fine}:

\begin{eqnarray}
s_+&=&s_x m_x+s_y m_y=\frac{\omega_{exc}^2}{\omega_{eff}^2}+\frac{\omega_{rf}^2}{\omega_{eff}^2}\cos(\omega_{eff}t)\\
s_-&=&s_y m_x - s_x m_y=\frac{\omega_{rf}}{\omega_{eff}}\sin(\omega_{eff}t)\\
s_z&=&\frac{\omega_{exc}\omega_{rf}}{\omega_{eff}^2}[1-\cos(\omega_{eff}t)]
\end{eqnarray}

\noindent with $\omega_{rf}$ the RF pulsation frequency, $\omega_{exc}=\frac{J_{exc}}{\hbar}$ the \textit{exchange} pulsation and $\omega_{eff}=\sqrt{\omega_{rf}^2+\omega_{exc}^2}$ the effective pulsation of the spin during its rotation.

In an homogeneous FM layer, the precession frequency due to the exchange interaction $\omega_{exc}\approx 10^{15}$~rad.s$^{-1}$ is very large compared to the RF frequency. This results in a small average component of the spin vector pumped along $z$: $s_z=\frac{\omega_{rf}}{\omega_{exc}}$, of the order of $10^{-5}$. However, this small spin rotation is counterbalanced by a large number of uncompensated spins due to the strong exchange and whose number equals $\mathcal{N}_{DOS}J_{exc}$ ($\mathcal{N}_{DOS}$ is the density of states). The total spin along the $z$ direction then writes $S_z\approx \mathcal{N}_{DOS}J_{exc}~s_z=\mathcal{N}_{DOS}\hbar\omega_{rf}$. One recovers the standard formula for the spin-current pumped at the ohmic FM/N interface if the interfacial spin-mixing conductance $g^{\uparrow\downarrow}$ is introduced hereafter. In the case of spin-pumping into evanescent states, the \textit{exchange} pulsation $\omega_{exc}$ can be of the order of magnitude of the RF pulsation or even smaller. To derive the average $s_z$ component pumped in a localized center, one has to perform a time average of $s_z$ on the carrier lifetime $\tau_n$ to give:

\begin{eqnarray}
s_z=\frac{\omega_{exc}\omega_{rf}}{\omega_{exc}^2+\omega_{rf}^2}\left[1-\frac{\sin(\omega_{eff}\tau_n)}{\omega_{eff}\tau_n}\right]
\end{eqnarray}

By analogy with the previous calculations relative to the bulk FM layer, and taking into account that the total number of uncompensated spins introduced by the \textit{tunneling} exchange interactions, $\mathcal{N}_{DOS}.\widetilde{J}_{exc}$, one can generalize the total spin accumulation ($\Delta \mu_z$) pumped along the $z$ direction as:

\begin{eqnarray}
\Delta \mu_z=\frac{(J_0 T)^2}{(J_0 T)^2+(\hbar \omega_{rf})^2}\left[1-sinc(\frac{\sqrt{(\hbar \omega_0)^2+(J_0 T)^2}}{\epsilon_n T})\right]\times \left[\hbar {\bf{m}}\times\frac{d {\bf{m}}}{dt}\right]
\end{eqnarray}

\noindent for any rotation $d{\bf{m}}/dt$ vector. It comes two important conditions on the effective exchange $\widetilde{J}_{exc}$ to generate significant spin-pumping at the FM/tunnel barrier/SC interface:

1) $\widetilde{J}$ must be larger than the intrinsic energy broadening $\Gamma$ (or equivalently $J_0>\epsilon_n$) corresponding to a time of interaction larger than the time of the spin precession.

2) $\widetilde{J}$ must be larger than the RF frequency energy $\hbar \omega_{rf}$ of the order of $\approx 40\mu eV$ in the present case. This condition corresponds to a characteristic spin precession time due to exchange, and necessary for any spin rotation, smaller than the magnetization ${\bf{m}}$ precession time itself.

Once these two conditions are satisfied, the spin-pumping effect at the FM/tunnel barrier/SC interface becomes efficient, a large rotation angle of the spin $s_z=\frac{\omega_{exc}\omega_{rf}}{\omega_{exc}^2+\omega_{rf}^2}$ compensating the small number of uncompensated spins $\mathcal{N}_{DOS}\widetilde{J}_{exc}$.

The total spin-current pumped ($I_s^p$) at the LS/SC channel interface, that is the \textit{source term}, equals $I_s^p=G_{sh}\Delta \mu_z$ where $G_{sh}$ is the Schottky conductance playing the role of the mixing conductance $g^{\uparrow \downarrow}$ for FM/N interfaces. We now proceed to the \textit{down-renormalization} of the spin-current pumped in the SC as described previously. In the light of the recent published works,\cite{bauer2012,vanwees2012} this total spin-current has then to be decomposed into the \textit{real} spin-current injected in the Ge channel added to a backflow of spin-current relaxing either into the localized states or into the FM reservoir by back-absorption. We have:

\begin{eqnarray}
I_s^p=G_{sh}\Delta \mu_z=G_{Ge}\Delta \mu_{Ge}+\frac{\Delta \mu_{LS}}{R_{LS}}\\
G_{sh}(\Delta \mu_{Ge}-\Delta \mu_{LS})=\frac{\Delta \mu_{LS}}{R_{LS}}\label{Gsh}
\end{eqnarray}

\noindent where $G_{Ge}=\tanh(t_N/l_{sf}^{Ge})/R_s^{Ge}$ is the spin conductance of the Ge layer of thickness $t_N$ and spin diffusion length $l_{sf}^{Ge}$ (Ref. \cite{vanwees2012}) and where $R_s^{Ge}=\rho_{Ge}\times l_{sf}^{Ge}$ is the corresponding bulk spin resistance. $\Delta \mu_{Ge}$ (\textit{resp.} $\Delta \mu_{LS}$) is the spin accumulation generated in the Ge layer (\textit{resp.} in the LS) and $R_{LS}=\tau_{sf}^{LS}/(e^2\mathcal{N}_{DOS}^{2D})$ is the spin resistance of the LS ($\tau_{sf}^{LS}$ is the corresponding spin lifetime). Eq. (\ref{Gsh}) describes the continuity of the spin-current backflow through the Schottky barrier. It results from these calculations that the effective spin-current $I_s^{Ge}$ injected in the Ge channel writes:

\begin{eqnarray}
I_s^{Ge}=\frac{1+R_{LS}G_{sh}}{\frac{1}{G_{Ge}}+\frac{1}{G_{sh}}+R_{LS}}\Delta \mu_z
\end{eqnarray}

with $\Delta \mu_z$ the spin accumulation generated in the localized states by spin-pumping like calculated previously. A zero Schottky conductance leads to zero spin-current. On the opposite case of a large Schottky conductance \textit{e.~g.} on increasing the temperature, the spin current pumped in the Ge channel writes $\frac{R_{LS}G_{Ge}}{1+R_{LS}G_{Ge}}\times G_{sh}\Delta \mu_z$ \textit{i.e.} it corresponds to the maximum spin-current pumped weighted by the ratio of spin-flips occurring in the channel itself over the total number of spin-flips also possible in the band of LS and parameterized by $1/R_{LS}$. Consequently, the \textit{real} spin current pumped into the Ge layer depends on the Schottky conductance and on the different spin-resistances involved in the spin-relaxation process. The main question that has to be addressed in the future is the fraction of the spin-current pumped and relaxing in the LS by spin-flip. Indeed, this part of the spin-current would contribute to the broadening of the FMR spectra but not to the ISHE voltage. Finally these parameters have to be determined experimentally in order to relate this microscopic model to our data.

\section{Conclusion}
We have demonstrated that at the FM/I/N interface where N is a
non-magnetic semiconductor channel we could inject a spin current by
spin pumping from the FM layer into the N channel at the
ferromagnetic resonance. We have also shown that the MgO tunnel
barrier is useful not only to overcome the conductivity mismatch
between CoFeB and Ge but also to keep the magnetic properties of the
FM after annealing the samples. There is no spin pumping nor inverse
spin Hall effect voltage signal on devices without barrier while it
clearly appears on the devices with the MgO barrier. Moreover there
is an enhancement of the ISHE signal and consequently of the spin
Hall angle when the device is annealed. A microscopic model involving interface states and \textit{evansecent} tunnel exchange coupling has been developed in order to explain spin pumping into Ge from a FM electrode through a tunnel barrier. We could finally find and discuss the spin Hall angle in $n$-Ge: $\sim 0.0011$ from the annealed sample and $\sim 0.00044$ from the as-grown one.

\section*{Acknowledgements}
J.-C. Rojas-Sanchez would like to acknowledge the Eurotalent CEA program for its financial support.

\appendix
\section{Angular dependence of the planar Hall effect and anomalous Hall effect\label{appendixA}}

\begin{figure}[h!]
\includegraphics[width=10 cm]{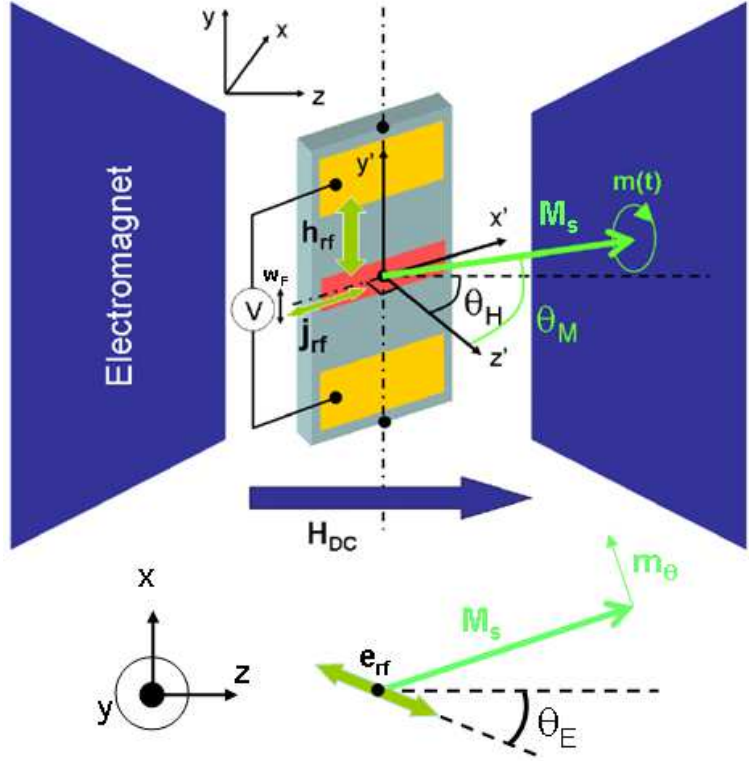}
\caption{Drawing of the experimental set-up along with the
definitions of the constants used in the calculations.} \label{fig_annex1}
\end{figure}

The sample is rotated in the electromagnet with the static magnetic
field $\vec{H}_{DC}$ applied perpendicular ($\theta_H$=0$^{\circ}$) to
parallel to the film plane ($\theta_H$=90$^{\circ}$). We define the
sample magnetization: $\vec{M}=\vec{M_{s}}+\vec{m}$ and the current
in the ferromagnetic electrode: $\vec{J}=\vec{J_{0}}+\vec{j}$ where
$\vec{M_{s}}$ is the equilibrium magnetization in static conditions.
It makes an angle $\theta_{M}$ with the normal to the sample
$\hat{z'}$. $\vec{m}(t)$ is the time-varying part of the
magnetization. We do not apply any bias current to the system:
$\vec{J_{0}}=\vec{0}$ and consider the RF current created in the
ferromagnetic layer by the non-vanishing RF electric field:
$\vec{j}=j_{0}cos(\omega t+\psi)\hat{x'}$ where $\psi$ is the
constant phase difference between the current and the magnetization
precession at resonance. The generalized Ohm's law then writes:\cite{Juretschke1960}

\begin{equation}
\vec{E}=\rho
\vec{j}+\frac{\Delta\rho}{M^{2}}(\vec{j}.\vec{M})\vec{M}-R_{H}\vec{j}\times\vec{M}
\end{equation}

\noindent $\rho$, $\Delta\rho$ and $R_H$ are the resistivity, the
anisotropic magnetoresistance and the anomalous Hall constant of the
ferromagnetic electrode. Since we experimentally measure a DC
voltage, we calculate the time average of $\vec{E}$:

\begin{equation}
\left\langle
\vec{E}\right\rangle=\frac{\Delta\rho}{M_s}^{2}\left[\left\langle
(\vec{j}.\vec{m}).\vec{M_s}\right\rangle+\left\langle
(\vec{j}.\vec{M_s}).\vec{m}\right\rangle\right]-R_H\left\langle
\vec{j}\times\vec{m}\right\rangle
\end{equation}

\noindent where: $\vec{M_s}=M_s sin(\theta_M -\theta_H)\hat{x}+M_s
cos(\theta_M -\theta_H)\hat{z}$, $\vec{m}=m_\theta cos(\theta_M
-\theta_H)\hat{x}+m_y\hat{y}-m_\theta sin(\theta_M
-\theta_H)\hat{z}$ and $\vec{j}=j cos\theta_H\hat{x}+j sin\theta_H
\hat{z}$. We then obtain the voltage:

\begin{equation}
V=\int_{-w_F/2}^{w_F/2}\left\langle E_y\right\rangle
dy=\frac{w_F\Delta\rho}{M_s}\left\langle jm_y\right\rangle sin\theta_M
-\frac{w_Fr_H}{M_s}\left\langle jm_\theta\right\rangle sin\theta_M
\end{equation}

\noindent where $r_H =M_s R_H$. The first term corresponds to the
planar Hall effect and the second one to the anomalous Hall effect.
$m_\theta$ and $m_y$ are then determined by solving the
Landau-Lifschitz-Gilbert (LLG) equation:

\begin{equation}
\frac{d\vec{M}}{dt}=-\gamma (\vec{M}\times\vec{H}_{eff})+\frac{\alpha}{M_s}\vec{M}\times\frac{d\vec{M}}{dt}
\end{equation}

\noindent where $\gamma$ is the gyromagnetic ratio, $\alpha$ the
damping factor and $\vec{H}_{eff}$ the effective magnetic field
given by:

\begin{equation}
\vec{H}_{eff}=H_{DC} \hat{z}+h_{rf} \hat{y}-4\pi (\vec{M}.\hat{z'}).\hat{z'}
\end{equation}

Here we only consider the shape anisotropy field. The radiofrequency magnetic field can be written as: $\vec{h}_{rf}=h_{rf0} cos(\omega t)\hat{y}$, $\omega=2\pi f$ where $f$=9.4 GHz is the X-band cavity frequency. In static conditions, the magnetization equilibrium angle $\theta_M$ is found by solving: $\vec{M}\times\vec{H}_{eff}=\vec{0}$ \textit{i.e.} $4\pi M_s sin(2\theta_M)-2H_{DC} sin(\theta_M -\theta_H)=0$. The resolution of the LLG equation yields:

\begin{eqnarray}
\dot{m_\theta}cos(\theta_M -\theta_H)&=&-\gamma Am_y+\gamma M_s h_{rf} cos(\theta_M -\theta_H)-\alpha \dot{m_y}cos(\theta_M -\theta_H) \\
\dot{m_y}&=&-\gamma Bm_\theta+\alpha \dot{m_\theta} \\
-\dot{m_\theta}sin(\theta_M -\theta_H)&=&-\gamma M_s h_{rf}
sin(\theta_M -\theta_H)+\gamma Cm_y +\alpha\dot{m_y}sin(\theta_M
-\theta_H)
\end{eqnarray}

\noindent where: $A=H_{DC} -4\pi M_s cos\theta_H cos\theta_M$,
$B=4\pi M_s cos(2\theta_M)-H_{DC} cos(\theta_M -\theta_H)$ and
$C=4\pi M_s sin\theta_H cos\theta_M$. By using:
$m_\theta=Re(\underline{m}_\theta e^{i\omega t})$,
$m_y=Re(\underline{m}_y e^{i\omega t})$ and
$h_{rf}=Re(h_{rf0}e^{i\omega t})$, we finally find:

\begin{eqnarray}
m_\theta &=&\frac{M_s h_{rf}cos(\theta_M -\theta_H)}{a^2+b^2}\left[a cos(\omega t)+b sin(\omega t)\right]\\
m_y &=&\frac{M_s h_{rf}cos(\theta_M
-\theta_H)}{a^2+b^2}\left[(\alpha
a+\frac{\gamma}{\omega}Bb)cos(\omega
t)-(\frac{\gamma}{\omega}Ba-\alpha b)sin(\omega t)\right]
\end{eqnarray}

\noindent where: $a=\alpha \left[A-Bcos(\theta_M -\theta_H)\right]$
and $b=(\gamma
/\omega)\left[AB+(\omega/\gamma)^2(1+\alpha^2)cos(\theta_M
-\theta_H)\right]$. The FMR spectrum is defined by $m_y$. Then,
after time averaging, we obtain:

\begin{eqnarray}\label{eqVPHE}
V_{PHE}&=&\frac{w_F\Delta\rho j_{0}h_{rf}cos(\theta_M -\theta_H)sin\theta_M}{2(a^2+b^2)} \nonumber \\
& &\hspace{4cm} \times \left[(\alpha a+\frac{\gamma}{\omega}Bb)cos\psi + (\frac{\gamma}{\omega}Ba-\alpha b)sin\psi\right] \\
V_{AHE}&=&\frac{w_Fr_H j_0 h_{rf}cos(\theta_M -\theta_H)sin\theta_M}{2(a^2+b^2)}\times \left[bsin\psi -acos\psi\right]
\end{eqnarray}

Here we point out that the rf electric field $\bold{e_{rf}}$ induces an
additional angular dependence because $j_0$ change with the DC magnetic field angle as
$sin(\theta_H+\theta_E)$ where $\theta_E$ is the direction
of the rf electric field in the cavity (see Fig.~\ref{fig_annex1}). 
%
Note that $j_0$ is proportional to the strength of the rf electric field, \textit{i.e.}
to the strength of the rf magnetic field. As a consequence the magnitude of
the electromotive forces $V_{PHE}$ and $V_{AHE}$ are
proportional to $h_{rf}^{2}$. Hence they exhibit a linear dependence
with the microwave power as shown in the main text. Such
linear dependence might allow to deduce either the ratio
$\Delta\rho/r_H$ or the phase shift $\psi$. Since the Hall
coefficient is of the order of $10^{-12}$ $\Omega cm/G$ and $M_s$ is
of the order of $10^3$ G, then $r_H=M_s R_H$ is much smaller than
the anisotropic magnetoresistance, $\Delta\rho$.\\
The OOP angular dependence of the symmetric component of either
$V_{PHE}$ or $V_{AHE}$ clearly shows a behavior different from that
of the ISHE out-of-plane angular dependence ($V_{ISHE}\propto
sin\theta_M$) as shown in Fig.~\ref{fig4oop}(c) where we considered
$\theta_E=-30^o$ and $\psi=15^o$.

\section{Charge backflow into the ferromagnet by the ISHE in Ge\label{AppendixB}}

\begin{figure}[h!]
\includegraphics[width=10cm]{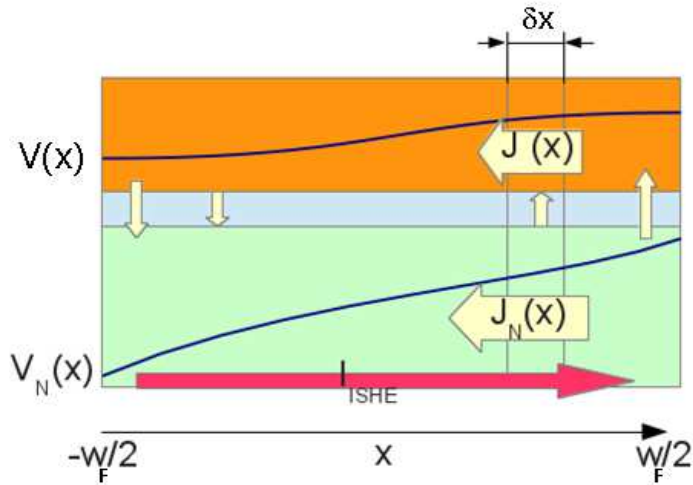}
\caption{Sketch of the CoFeB/MgO/Ge trilayer with the currents and potentials.}
\label{figBackflow}
\end{figure}

 At the ferromagnetic resonance of the CoFeB electrode, the
combination of spin pumping and ISHE creates a charge current (in A/m)
$I_{ISHE}$ in the Ge layer. Part of this charge current flows back
to the ferromagnetic and tantalum capping layers which affects the
estimation of $V_{ISHE}$ and $\theta_{SHE}$ in germanium. In the
following, we make an estimation of this backflow current. In
Fig.~\ref{figBackflow}, the current density crossing the interface
at $x$ corresponds to the variation of the current in the layers:

\begin{eqnarray}\label{AppendixB_1}
t\delta j(x)=\frac{V_N-V}{RA}\delta x\\
t_N\delta j_N(x)=\frac{V-V_N}{RA}\delta x
\end{eqnarray}

where $t$, $j$ and $V$ are the thickness, current density and potential in the Ta/CoFeB bilayer; $RA$ is the resistance-area product of the interface between CoFeB/MgO and Ge. The current densities in each layer with conductivities $\sigma$ and $\sigma_N$ can be written:

\begin{eqnarray}\label{AppendixB_2}
j(x)=-\sigma\partial_xV(x)\\
j_N(x)=-\sigma_N\partial_xV_N(x)
\end{eqnarray}

the current conservation involving the current source due to spin pumping and ISHE gives:

\begin{equation}
tj(x)+t_Nj_N(x)=-I_{ISHE}
\end{equation}

which can also be written:

\begin{equation}
t\sigma\partial_xV(x)+t_N\sigma_N\partial_xV_N(x)=I_{ISHE}
\end{equation}

by using the symmetry of the system, we set the origin of $x$ in the middle of the trilayer and find:

\begin{equation}\label{AppendixB_3}
t\sigma V(x)+t_N\sigma_NV_N(x)=I_{ISHE}x
\end{equation}

Using Eq.~\ref{AppendixB_1} and ~\ref{AppendixB_2}, we can write:

\begin{equation}
\partial_{x}^{2}\left(V(x)-V_N(x)\right)=\left(\frac{1}{t\sigma}+\frac{1}{t_N\sigma_N}\right)\frac{V(x)-V_N(x)}{RA}
\end{equation}

which gives the following solution:

\begin{equation}\label{AppendixB_4}
V(x)-V_N(x)=asinh\left(\frac{x}{\lambda}\right)
\end{equation}

with:

\begin{equation}
\left(\frac{1}{\lambda}\right)^{2}=\left(\frac{1}{t\sigma}+\frac{1}{t_N\sigma_N}\right)\frac{1}{RA}
\end{equation}

combining Eq.~\ref{AppendixB_3} and ~\ref{AppendixB_4} yields the potentials:

\begin{eqnarray}
V(x)=\frac{I_{ISHE}}{t\sigma +t_N\sigma_N}x-a sinh\left(\frac{x}{\lambda}\right)\frac{t_N\sigma_N}{t\sigma+t_N\sigma_N}\\
V_N(x)=\frac{I_{ISHE}}{t\sigma +t_N\sigma_N}x+a sinh\left(\frac{x}{\lambda}\right)\frac{t\sigma}{t\sigma+t_N\sigma_N}
\end{eqnarray}

The current in the Ta/CoFeB bilayer (proportional to the derivative of $V$) vanishes at the edges ($x=\pm w_F/2$) which gives access to the constant $a$:

\begin{equation}
a=\frac{\lambda I_{ISHE}}{t_N\sigma_N cosh((w_F/2\lambda)}
\end{equation}

Then the ratio between the induced voltage $U$ in Ge and the current $I_{ISHE}$ is given by:

\begin{equation}
\frac{U}{I_{ISHE}}=\frac{2V_N(w_F/2)}{I_{ISHE}}=\frac{w_F}{t\sigma+t_N\sigma_N}\left[1+\frac{t\sigma}{t_N\sigma_N}\frac{2\lambda}{w_F}tanh\left(\frac{w_F}{2\lambda}\right)\right]
\end{equation}

where: $t\sigma=t_F\sigma_F+t_{Ta}\sigma_{Ta}$. $t_{Ta}$ (resp. $\sigma_{Ta}$) is the thickness (resp. conductivity) of the tantalum capping layer.

\end{document}